\documentclass[times]{aastex62}

\newcommand{\rxte}{\textit{RXTE}}

\shorttitle{Collapse of a thick disk due to a Type I X-ray burst}
\shortauthors{Fragile et al.}
\begin{document}

\title{Simulating the Collapse of a Thick Accretion Disk due to a Type I X-ray Burst from a Neutron Star}

\author[0000-0002-5786-186X]{P. Chris Fragile}
\affiliation{Department of Physics \& Astronomy, College of Charleston, 66 George St., Charleston, SC, USA 29424}
\email{fragilep@cofc.edu}

\author[0000-0001-8128-6976]{David R. Ballantyne}
\affiliation{Center for Relativistic Astrophysics, School of Physics, Georgia Institute of Technology, 837 State Street, Atlanta, GA, USA 30332}
\email{david.ballantyne@physics.gatech.edu}

\author{Thomas J. Maccarone}
\affiliation{Department of Physics \& Astronomy, Texas Tech University, Box 41051, Lubbock, TX 79409-1051, USA}

\author{Jason W. L. Witry}
\affiliation{Department of Physics \& Astronomy, College of Charleston, 66 George St., Charleston, SC, USA 29424}



\begin{abstract}
We use two-dimensional, general relativistic, viscous, radiation hydrodynamic simulations to study the impact of a Type I X-ray burst on a hot and geometrically thick accretion disk surrounding an unmagnetized, non-rotating neutron star. The disk is initially consistent with a system in its low/hard spectral state, and is subject to a burst which rises to a peak luminosity of $10^{38}$~erg~s$^{-1}$ in $2.05$~s. At the peak of the burst, the temperature of the disk has dropped by more than three orders of magnitude and its scale height has gone down by more than one order of magnitude. The simulations show that these effects predominantly happen due to Compton cooling of the hot plasma, and clearly illustrate the potential cooling effects of bursts on accretion disk coronae. In addition, we demonstrate the presence of Poynting-Robertson drag, though it only enhances the mass accretion rate onto the neutron star by a factor of $\sim 3$--$4$ compared to a simulation with no burst. Simulations such as these are important for building a general understanding of the response of an accretion disk to an intense X-ray impulse, which, in turn, will be crucial for deciphering burst spectra. Detailed analysis of such spectra offers the potential to measure neutron star radii, and hence constrain the neutron star equation of state, but only if the contributions coming from the impacted disk and its associated corona can be understood. 
\end{abstract}


\section{Introduction} \label{sec:intro}
Type I X-ray bursts result from unstable thermonuclear burning of accreted matter on the surface of a neutron star \citep[e.g.,][]{Strohmayer2006,gk18}. The sudden release of energy is thermalized by the neutron star atmosphere and radiated away as X-rays over a timescale of several seconds to minutes \citep[e.g.,][]{belian76,lewin93,galloway08}. During its peak, the luminosity of the burst significantly outshines the surrounding accretion disk, and may even exceed the local Eddington luminosity of the star \citep[e.g.,][]{hansen75,lewin84,tawara84,galloway08}. Following the burst, material resumes accumulating on the stellar surface, leading to subsequent X-ray bursts with a recurrence time of hours to days, depending on the metallicity and accretion rate of the infalling gas \citep[e.g.,][]{fujimoto81,bildsten98}. As the neutron star surface is the origin of the observed X-rays, there is significant interest in using the spectra and luminosities of X-ray bursts to measure the physical size of the star, a measurement necessary to constrain the equation of state of nuclear matter \citep[e.g.,][]{guver12,kajava14,nattila17}. 

Between each X-ray burst cycle the neutron star system is characterized as an accretion-powered, low-mass X-ray binary, and appears in one of a number of different X-ray spectral and timing states, which are likely related to the geometry of the accretion flow onto the neutron star \citep[e.g.,][]{hvk89,shirey99}. The accretion disk and its accompanying corona are therefore additional X-ray sources that are present during a burst event and must be taken into account when X-ray burst data is utilized to measure neutron star radii. In a traditional X-ray burst analysis, the X-ray spectrum produced by the disk and corona before the burst (also known as the `persistent' spectrum) is used as the background spectrum to be subtracted from the burst spectrum \citep[e.g.,][]{galloway08}. The implicit assumption behind this procedure is that the disk and corona are not affected by the burst and their contribution to the spectrum does not change. However, recent analyses of both a large number of bursts in the \rxte\ archive \citep{worpel13,worpel15} and two long individual bursts \citep{Zand2013,Keek2014sb1} show strong evidence that the normalization of the persistent spectrum may be elevated during the burst by factors of $\sim 1$--$10$, potentially indicating an increase in the accretion rate, possibly due to Poynting-Robertson (PR) drag \citep{Blumenthal74,Walker89,Walker1992}. In addition, careful examination of a small number of bursts has shown that the high-energy ($>30$~keV) persistent emission appears to drop during the burst \citep{mc03,chen13,ji13,kajava17,chen18}, perhaps due to increased Compton cooling in the corona. Taken together, both of these results suggest that the accretion disk and corona may be significantly affected by the strong photon field during an X-ray burst \citep{degenaar18}. 

The physical response of an accretion disk and corona to the impact of an X-ray burst is not currently well understood. As discussed by \citet{be05}, a wide range of processes and responses are possible, including outflows (driven by radiation pressure), inflows (driven by PR drag), and heating and cooling effects. This situation must be improved if X-ray bursts are to be exploited as effective probes of neutron star physics. Toward this end, this paper presents the first numerical simulation of an accretion disk subject to the sudden, intense radiation field of an X-ray burst. The results clearly show the impact of the burst on the disk and should provide important guidance for correctly including the effects of the disk when modeling burst spectra. The next section describes the numerical setup, with the results described in Section~\ref{sec:results}. A short summary and discussion are presented in Section~\ref{sec:concl}.

\section{Numerical Simulations} \label{sec:methods}

The simulations presented in this paper use the \textit{Cosmos++} computational astrophysics code \citep{Anninos05,Fragile12,Fragile14}. The numerical treatments of the hydrodynamics, radiation, and viscosity are nearly identical to those described by \citet{Fragile18}, so we forgo providing extensive details here. The simulations are two-dimensional, axisymmetric, and cover a radial range from $r_\mathrm{min} = 10.7$ km to $r_\mathrm{max} = 1531$ km, with exponential spacing. We use ``outflow'' boundaries at both the inner and outer radial boundaries, which effectively means we are assuming any boundary layer effects happen inside $r_\mathrm{min}$. The full range of $\theta$ is covered with a latitude coordinate, $x_2$, related to $\theta$ by $\theta = x_2 + 0.25 \sin (2x_2)$, which concentrates resolution toward the midplane. A layer of static mesh refinement further enhances the resolution in the region of interest. The base resolution is $96^2$, with the additional layer of refinement covering the region $r_\mathrm{min} \le r \le 535$ km and $0.1\pi \le x_2 \le 0.9\pi$ ($27.5^\circ \le \theta \le 152.5^\circ$) for an effective peak resolution of $192^2$ over the disk.

Both the X-ray burst properties \citep{gk18} and the strength of the burst-disk interaction \citep{ji14} are observed to depend on whether the system is in the low/hard or high/soft state when the burst is ignited. These two states are thought to correspond to different accretion regimes and disk geometries. In the low/hard state, the accretion rate is relatively low and the thin accretion disk is possibly truncated at small radii into a geometrically thick, hot flow that acts as a large corona \citep[e.g.,][]{done07}. In contrast, the high/soft state is related to more rapid accretion through an untruncated, geometrically thin disk extending down to the stellar surface. In this paper, a geometrically thick disk is assumed for the simulation, which would correspond to the system being in the low/hard state. The results for a thin disk in the high/soft state will be presented in subsequent work.

In constructing the thick disk, we have assumed the spacetime around our $1.45 M_\odot$ neutron star can be adequately described by the Schwarzschild metric (in Kerr-Schild polar coordinates), appropriate for non- or slowly rotating neutron stars \citep[e.g.][]{Hartle67}. We then construct the thick disk as an initially stationary hydrodynamic torus, of the type first described by \citet{Kozlowski78}. We choose a constant specific angular momentum profile, with the inner edge of the torus initially at $r_\mathrm{in} = 47$ km and its center at $r_\mathrm{cen} = 86$ km. This fully specifies the torus solution, other than an arbitrary density normalization, which we adjust to give us the desired mass accretion rate. Accretion is driven by a \citet{Shakura73} type $\alpha$-viscosity with the viscosity coefficient calculated following standard disk theory as $\mu=\nu\rho=\alpha\rho c_s H$, where $\alpha = 0.025$, $c_s = \sqrt{P_\mathrm{tot}/\rho}$ is the thermal sound speed, $P_\mathrm{tot} = P_\mathrm{gas} + P_\mathrm{rad}$ is the total pressure, $H = c_s/V^\phi$ is the disk height, and $V^\phi$ is the azimuthal coordinate velocity of the gas. This viscosity rapidly (on the dynamical timescale) redistributes the angular momentum of the initial torus into a nearly Keplerian profile.

Although this torus setup is fairly standard in accretion simulation literature, it does present two related disadvantages for our current purposes: 1) Because the torus is isolated, mass is not replenished as it leaves the simulation domain, either as a result of accretion through the inner boundary (i.e., onto the neutron star) or expulsion beyond the outer boundary (i.e., as part of a wind). 2) This also means that the mass accretion rate through the inner boundary is never truly steady, instead rising sharply at early times and then gradually decaying as the simulation proceeds. For our current setup, the mass accretion rate peaks around $\dot{m} = \dot{M} c^2/L_\mathrm{Edd} = 10^{-3}$, where $L_\mathrm{Edd} = 1.7 \times 10^{38}$ erg s$^{-1}$, but decays to below $10^{-5}$ by the end of the simulation. This leads to a net loss of mass in the domain of $>97$\% over the first 2.25 s of the simulation. This dramatic loss of mass is one reason why these simulations were not run for the entire duration of the burst.

X-ray bursts are typically not observed at such low accretion rates \citep{gk18}, but numerically simulated disks at more realistic mass accretion rates tend to spontaneously become thin \citep[see, e.g., model B in][]{Ohsuga11}, in contrast with what appears to be the case in nature. This is an issue that has yet to be resolved. Since our interest in this Letter is to see what happens to a disk in the low/hard state during an X-ray burst, we make the choice to use an accretion rate that we know will produce a geometrically thick (radiatively inefficient) disk. We suspect our results would still hold at higher $\dot{m}$ (provided the disk is still in the low/hard state), but this will need to be checked.

Since most Type I X-ray bursts exhibit a fast rise and power-law or exponential decay \citep{lewin93}, we model the burst luminosity, $L(t)$, using the so-called Norris model from the Gamma-Ray Burst (GRB) community \citep{Norris05}:
\begin{equation}
L(t) = L_0 e^{2(\tau_1/\tau_2)^{1/2}} e^{\frac{-\tau_1}{t-t_s} - \frac{t-t_s}{\tau_2}} ~,
\label{eqn:Norris}
\end{equation}
where $L_0$ is the peak burst luminosity, $t_s$ is the burst start time, and $\tau_1$ and $\tau_2$ characterize the burst rise and decay. In this work we choose $L_0 = 10^{38}$~erg~s$^{-1}$, $t_s = -0.4$~s, $\tau_1 = 6$~s, and $\tau_2 = 1$~s, which produces a burst that peaks at a simulation time of $t=2.05$~s, lasts for about 10~s, and has a total energy output of $3\times 10^{38}$~erg. The simulation duration (2.25~s) and burst peak (2.05~s) times are much longer than the dynamical and thermal timescales in the inner parts of the disk ($t_\mathrm{dyn} = 2\pi/\Omega \sim 0.001$~s and $t_\mathrm{th} = 2\pi/(\alpha \Omega) \sim 0.05$~s, respectively), and comparable to the viscous timescale ($t_\mathrm{vis} = r^2/(\alpha c_s H) \sim 10$~s).

The burst luminosity is introduced into the simulation domain as an outward pointing radiation flux through the inner radial boundary, $R^r_t (t, r_\mathrm{min}) = -L(t)/(4\pi r_\mathrm{min}^2)$. Any contribution to this luminosity coming from accretion onto the neutron star surface is neglected, as the accretion luminosity from this disk is quite low ($< 10^{33}$ erg s$^{-1}$). We also ignore the angular velocity of the emitting material, consistent with our use of the Schwarzschild metric. Prograde (in the same direction as the orbital motion of the disk) rotation of the neutron star would cause PR drag to be less effective. For the fastest rotating neutron stars, with spin periods of a few ms, the observable effects of PR drag are expected to virtually disappear \citep{Walker1992}. On the other hand, while the $\mathbf{M}_1$ closure scheme we use for the radiation correctly captures PR drag in the limit of a distant point source \citep[e.g.,][]{Wielgus16}, for a finite-size source, such as a neutron star, because $\mathbf{M}_1$ does not properly handle light coming from extended sources, it underestimates the drag \citep[see Fig. 1 of][]{Oh11}. It remains for more accurate radiation treatments to fully assess the quantitative impact of PR drag during Type-I X-ray bursts.

As we will show, the dominant effect of the burst radiation comes from Compton cooling of the hot, thick disk. The strength of the Compton cooling goes as $4\rho \kappa^\mathrm{s}(T_\mathrm{gas}-T_\mathrm{rad})/m_e$, where $\rho$ is the gas density, $\kappa^\mathrm{s} = 0.34$ cm$^2$ g$^{-1}$ is the electron-scattering opacity, $m_e$ is the mass of the electron, and $T_\mathrm{gas}$ and $T_\mathrm{rad}$ are the gas and radiation temperatures, respectively. Typical values are $T_\mathrm{gas} \sim 10^{10}$ K (for disk plasma prior to the burst) and $T_\mathrm{rad} \approx 4 \times 10^6 ~\textrm{K} = 0.3$ keV (near the inner radial boundary at the peak of the burst).

\section{Results} \label{sec:results}

\begin{figure}
\includegraphics[width=\textwidth]{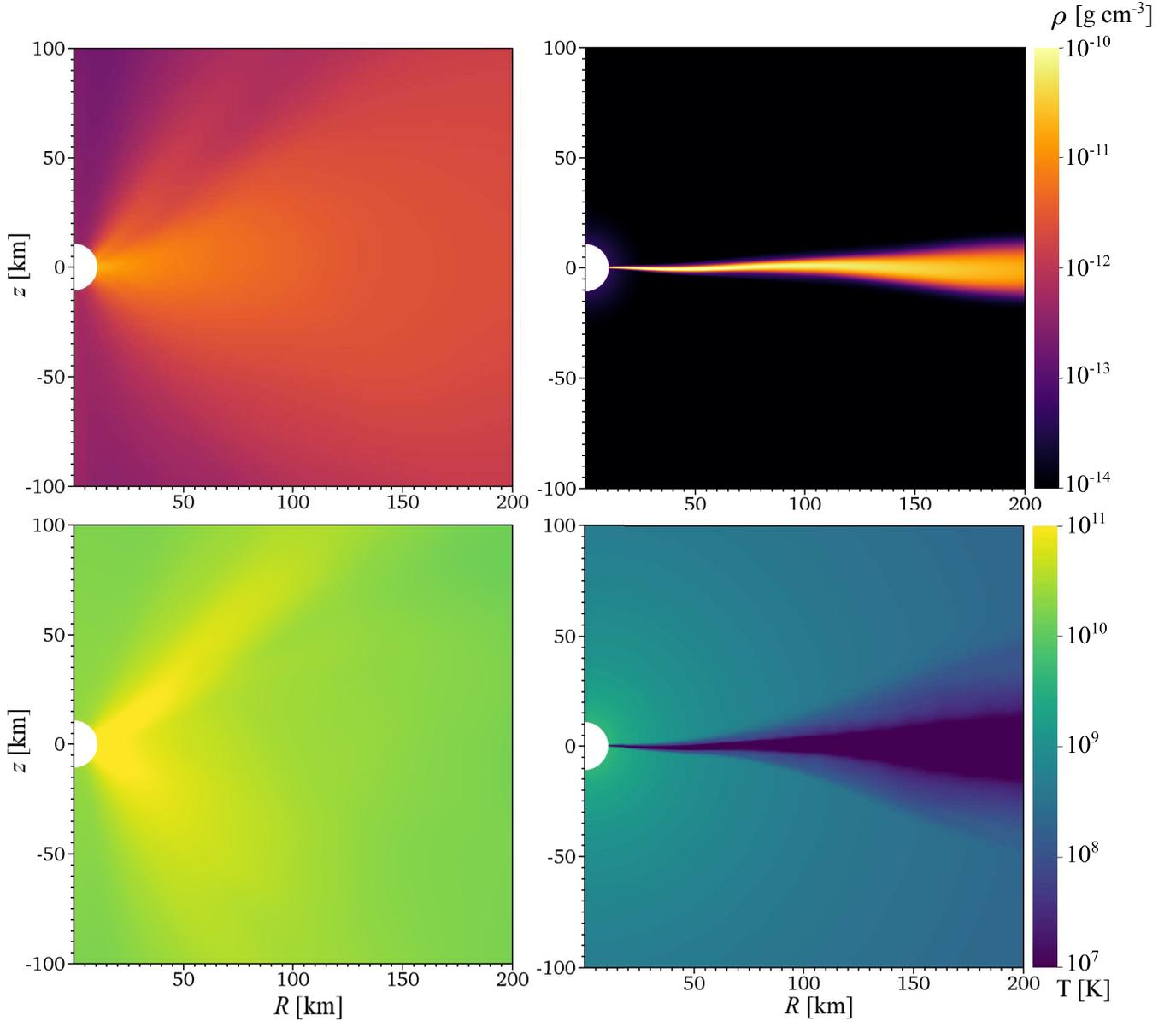}
\caption{Logarithm of the mass density (in g~cm$^{-3}$, top) and temperature (in K, bottom) from simulations without (left) and with (right) a Type I X-ray burst. All data are time-averaged over the final 0.5~s of each simulation ($1.75\,\mathrm{s} \le t \le 2.25\,\mathrm{s}$). The X-ray burst causes the initially hot and thick accretion flow to collapses into a disk that is both thinner and cooler by more than an order of magnitude.}
\label{fig:density}
\end{figure}

To establish the importance of the X-ray burst on the evolution of our accretion disk, two otherwise identical simulations were run, one with the burst activated, the other without. Figure \ref{fig:density} shows the dramatic contrast in density structure and temperature between the two simulated accretion disks when averaged over the final $0.5$~s (close to the peak of the burst). The disk without the burst (left) is hot ($T \gtrsim 10^{10}$~K)\footnote{N.B.: Pair production, if included in the code, would likely limit the maximum temperature to $T \lesssim 10^{10}$~K.} and geometrically thick ($H/R \approx 0.3$), as expected based on the initialization of the simulation. The case with the burst (right) is almost diametrically opposite: cold ($T \lesssim 10^7$~K) and geometrically thin ($H/R \lesssim 0.01$)\footnote{It is worth mentioning that the disk remains effectively optically thin throughout the simulation; however, this might not be true at higher mass accretion rates.}. Clearly, the burst radiation had a conspicuous effect on the structure of the accretion flow.

In this low accretion rate, thick-disk limit, we expect the dominant physical effect of the burst to come from inverse Compton cooling of the disk plasma by the burst photons. \citet{degenaar18} shows that this should begin to happen once the burst luminosity becomes roughly equal to the accretion luminosity, which is almost immediately in our simulations. The added cooling can be enough to dramatically lower the temperature and remove much of the thermal pressure support in the disk, resulting in its vertical collapse, consistent with what is seen in Figure \ref{fig:density}. As further proof, we ran another simulation that included a burst but with the Compton cooling term turned off in our radiation package. The results of that simulation, with a disk temperature and scale height within a factor of two of the no-burst simulation, confirm the principal role of Compton cooling. 

Another possible impact of the burst on the disk could come through radiation pressure driving material away. However, we find that at no time is more than 4\% of the total mass in our simulation domain caught up in an unbound ($u_0 < -1$) outflow ($v^r > 0$), i.e. a wind. Further, that 4\% is a transient phase early in the simulation ($t < 0.1$ s), well before the burst peak; at later times, the mass caught up in an unbound outflow becomes completely negligible. The mass in bound, but outflowing, material is much larger ($\gtrsim 50$\%), but this is to be expected as the initial torus is viscously spreading (about half of the material moving outward and half moving inward). Thus, we conclude that the differences we see in Figure \ref{fig:density} can not be attributed to radiation pressure driving, consistent with the sub-Eddington nature of our burst. 

\begin{figure}
\includegraphics[width=0.5\textwidth]{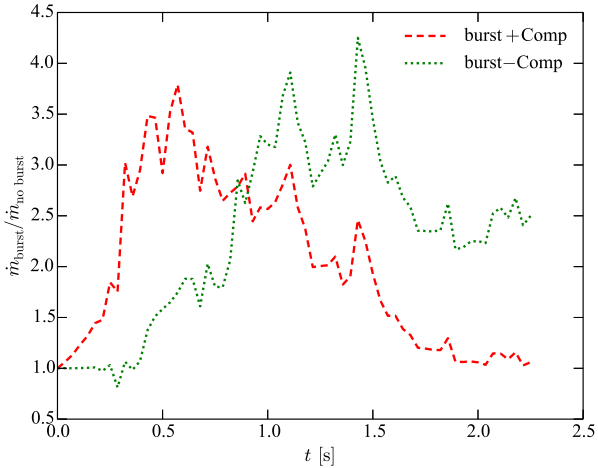}
\caption{Ratio between the mass accretion rate onto the neutron star for simulations with and without a Type I X-ray burst. The two curves distinguish between burst simulations where Compton cooling is included (red, dashed) and where it is turned off (green, dotted). Enhanced accretion of approximately equal magnitude occurs during both simulations, but the effects of Compton cooling boost the accretion rate earlier than PR drag alone.}
\label{fig:massflux}
\end{figure}

\begin{figure}
\includegraphics[width=0.5\textwidth]{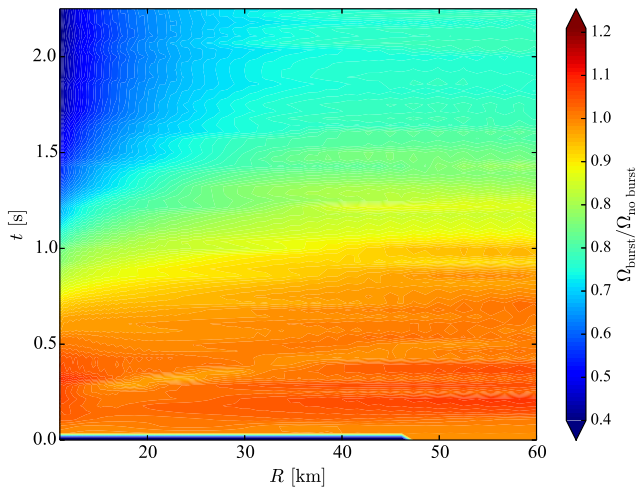}
\caption{Spacetime diagram of the ratio between the density-weighted orbital angular frequency for simulations with and without a Type I X-ray burst. The orbital frequencies of the two disks are nearly identical until $\approx 0.75$~s when the simulation with the burst begins to orbit at less than the Keplerian frequency. The ratio continues to decline all through the disk as the burst luminosity grows. This result indicates an additional torque is acting on the gas that is growing with the burst luminosity, exactly the hallmark of PR drag.}
\label{fig:Omega}
\end{figure}

A third predicted effect of Type I X-ray burst photons on the surrounding accretion disk is to drive additional accretion through PR drag. The red-dashed line in Figure \ref{fig:massflux} shows that the X-ray burst does enhance the mass accretion rate through the inner simulation boundary by factors of $\approx 3$--$4$, but it is obvious that the enhancement to $\dot{m}$ does not happen on the same timescale as the burst (which peaks at 2.05 s). Before addressing this discrepancy, let us first look for direct evidence of PR drag in the simulations. Figure \ref{fig:Omega} shows a spacetime diagram of the ratio between the density-weighted orbital angular frequency of gas in the disk for the simulations with and without a burst. In the case without a burst, the orbital angular frequency maintains a steady Keplerian profile throughout the simulation (not shown). However, this ratio plot shows that the simulation with a burst has orbital angular frequencies that begin to decay (orbit slower) on exactly the timescale of the burst, as would be expected if the disk material were being acted upon by an additional torque. 

However, this PR drag can not be the only cause for the enhanced accretion seen in our burst + Compton simulation (Figure \ref{fig:massflux}; red, dashed curve). First off, the enhanced accretion starts almost immediately and peaks well before Figure \ref{fig:Omega} shows significant deviations in the angular momentum profile ($t\gtrsim 0.75$~s). Instead, this early accretion appears to be tied to the Compton-driven cooling and collapse of the disk, which happens on the local thermal timescale. As evidence of this link, note that the simulation with a burst but without Comptonization (Figure \ref{fig:massflux}; green, dotted curve) does not show the same early enhancement of accretion. In both cases, though, the accretion rate peaks before the burst does. This has to do with the fact that the inner part of the disk can not refill (by accreting material from larger radii) as fast as it is being drained by these additional processes. The mass accretion rate onto a neutron star during a burst will, thus, be sensitive to both the strength of the PR drag and the time-varying surface density profile of the disk. 

\section{Discussion \& Summary} \label{sec:concl}

Using general relativistic, viscous, radiation hydrodynamic simulations, we have confirmed that the emission from a Type-I X-ray burst can affect a surrounding hot, thick accretion flow in at least two important ways: 1) the burst can Compton cool it by orders of magnitude, resulting in a temporary transition to a much cooler, thinner disk; and 2) the combination of the collapse of the disk and Poynting-Robertson drag can amplify the mass accretion rate by a factor of a few during the burst. This is the first time these effects have been demonstrated in such simulations.%, and illustrate the importance of considering the influence of the burst on the underlying accretion flow when analyzing X-ray burst observations.

This study focused exclusively on Type I X-ray bursts occurring during the low/hard state. We find that the relative thickness of the disk in this state presents a conveniently large target for the burst photons. The rapid ($t \lesssim \mathrm{a~few}~t_\mathrm{th}$) cooling in the simulations is in good agreement with observations of a prompt drop in the high energy ($> 30$~keV) flux during real X-ray bursts in the hard state \citep[e.g.,][]{chen18}. The three order of magnitude drop in temperature, however, is significantly larger than what is inferred from observations. Part of this discrepancy can be explained by the fact that we did not include the effects of pair production, which would have acted as a thermostat, regulating the maximum temperature of the plasma. Additionally, the unrealistically low disk mass accretion rate we considered probably contributed to the excessively high temperatures.

Another highlight of our simulations was the confirmation of PR drag as an important process acting on an accretion disk during a burst (Figure \ref{fig:Omega}). However, its impact was relatively modest, as it only enhanced the mass accretion rate onto the neutron star by factors of $\approx 3$--$4$, less than the observed increases in the persistent flux of $\ga 4$ \citep{worpel15}. However, we have to be careful here. Although we have chosen the most ideal scenario for PR drag, in that the applied radiation field is from a non-rotating neutron star, the $\mathbf{M}_1$ closure scheme employed underestimates PR drag from extended sources. Therefore, future simulations will be needed to better quantify the effect of PR drag.

In summary, this paper presents the first simulation of an accretion disk impacted by a powerful X-ray burst. The parameters were chosen so that the effects of the burst luminosity and PR drag would be maximized. This was sufficient to demonstrate that a Type I X-ray burst going off in the hard state may lead to important structural and thermodynamic changes in an accretion disk. These changes need to be understood before, for example, burst spectroscopy could be used to measure neutron star radii.

\acknowledgments

P.C.F. and J.W.L.W. acknowledge support from SC NASA EPSCoR RGP 2017 and National Science Foundation grant AST-1616185. This work used the Extreme Science and Engineering Discovery Environment (XSEDE), which is supported by National Science Foundation grant number ACI-1053575. D.R.B. and T.J.M. thank the International Space Science Institute, where part of this work was carried out, for their hospitality.

%\bibliography{refs}

\bibliographystyle{aasjournal}


\end{document}